# The egg steamer paradox

*S. Flach*
*Center for Theoretical Physics of Complex Systems, Institute for Basic Science (IBS), Expo-ro 55, Yuseong-gu, Daejeon 34126, South Korea*

*S. Parnovsky*
*National Taras Shevchenko University of Kyiv, Observatorna St, 3, Kyiv, 02000, Ukraine*

*A.A. Varlamov*
*CNR-SPIN, Via del Cavaliere del Fosse, 00133 Rome, Italy*

## *Abstract*

*Why do we need to pour less water in an egg steamer to prepare more eggs to the same degree of "doneness"? We discuss the physical processes at work in the electric egg steamer and resolve this seeming paradox. We demonstrate that the main heat transfer mechanism from steam to egg is due to latent heat through condensation. This not only explains the paradox, but also allows us to estimate the amount of water reduction. Comparing the preparation time to the one for traditional boiling, we estimate the eggshell temperature during steaming. We also describe the device design and provide further theoretical estimates and experimental kitchen measurement data for this appealing kitchen experiment that can be easily accomplished at home.*

New gadgets invade our kitchens at an ever quickening pace. Among the earliest automated kitchen gadgets is the electric egg cooker, also called an egg boiler. The term 'egg steamer' would suit better, since the eggs are steamed rather than boiled. Yet, the outcome is exactly the same as those familiar for a boiled egg.

**The device design and operation**

The typical egg cooker is shown in Fig. 1. It consists of a 300-400 W electric heater (we used a 350 W cooker for experiments), bowl-shaped heating plate designed to hold a small amount of water, the lid with a few small holes, and the egg holder. Crucially, the latter is placed above the heating plate and the water in it is placed above the heating plate and can host up to seven eggs (which are placed in the separate cells).

The cooking starts with pouring a relatively small amount of water into the measuring cup. This amount depends on the number of eggs, and the desired final degree of their "doneness": soft-, medium- or hard-boiled. The measuring cup is then emptied onto the heating plate. Note that the eggs remain positioned above the water level, without any direct contact with it, throughout the cooking process. The transparent lid is placed on top and covers the entire device. There are small holes in the lid order to allow for steam to escape. The number of openings can vary from one to three depending on the manufacturer's design, and their diameter is of the order of 1 cm. The automatic switch off of the device is set in action once all water is evaporated, and the temperature of the heating plate exceeds the established limit. Simpler versions only produce a loud beep signal, leaving the shutting off task to the kitchen master.

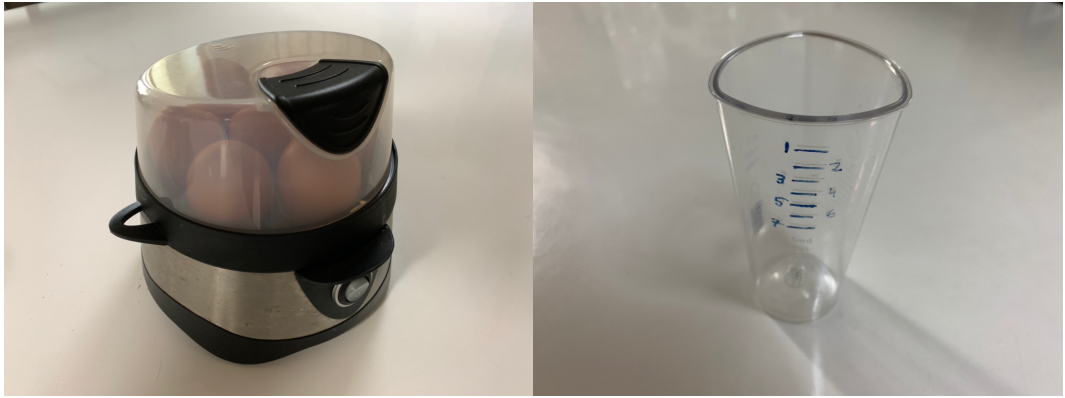

*Fig. 1 Electric egg steamer. The measuring water cup is included in the steamer set.*

The principle of operation of the device appears to be simple. Once switched on, the water heats up and after a certain amount of time starts to boil, producing steam. Hot steam rises and heats the eggs up. Some amount of steam continuously escapes through the openings in the lid, tending to equalize the inner and outer pressures.

When all is done in accordance with the above instructions, the eggs are cooked to the desired degree. The rough steamer preparation times $T_s$ are about 7-8 minutes for soft boiled eggs, and 16-20 minutes for hard boiled eggs (see the discussion below). In comparison, when boiling eggs in a pot in a conventional manner, the time for soft boiled eggs amounts to 4-5 minutes, and for hard boiled eggs to 10-14 minutes.

**The paradox**

The measuring cup in Fig.1 shows that less water is required for more eggs to be steamed to the same state (e.g. soft-boiled). This reciprocal relation ("more requires less") looks surprising, and one might even erroneously conclude that it needs zero water for infinite egg number! Nevertheless, astonishing as it might seem, the reciprocal law (less water for more eggs) certainly works well for

n=1,…,7 eggs. In order to explain that fact, one has to recall the known mechanisms of heat transfer.

**Heat transfer mechanisms**

Textbooks usually describe the three most common heat transfer mechanisms.
The first one is thermal conductivity i.e., the transfer of heat from the warmer areas to the colder ones through the thermal motion of molecules, without any mass flow. Thermal conductivity is the main mechanism for transferring heat through a solid, in which the atoms or the molecules cannot move far from their equilibrium positions. In the case of traditional boiling of an egg in water, this is the obvious mechanism to transfer energy from water molecules to the eggshell through inelastic collisions.
How effective is this mechanism in the egg steaming process? The steam in the device at standard atmosphere pressure has roughly the same temperature as boiling water: *100°C* which density here is *958 kg/m$^3$*. What concerns the density of steam at these conditions it is *0.6 kg/m$^3$* - almost *1600* times smaller. Hence, transfer of the equivalent amount of energy would take *1600* times longer when steaming as compared to boiling. Since the egg in the device is cooked only twice slower than in boiling water, the heat conductivity through the vapour can be excluded from the candidates for heat supply.

The second heat transfer mechanism is convection. It occurs due to the flows in the media. We exclude this mechanism for the egg steamer. Indeed, in order to reduce the 100 hours needed for egg preparation with heat conductivity down to the observed 10 minutes, we need to generate powerful currents with flow velocities which are expected to rupture the lid off. Since the plastic lid is stable, we can exclude convection.

The third mechanism is heat radiation. It is responsible for culinary wonders such as baking of the Neapolitan pizza in a wood oven [1]. Nevertheless, it is completely ineffective in the case of an egg cooker due to the relatively low temperatures of all its parts.

Yet, the release of the latent heat also occurs in the process of the vapour condensation which is nothing else as the phase transition of the first order. Thus, this fourth mechanism transfers provides the heat transfer from the heater to the colder surfaces in the process of steam condensation. The molecules of the water get enough energy during the heating and the process of evaporation to break bonds with neighbouring molecules in the liquid. After that, they escape into the gaseous steam phase. It is this energy which is later released during their re-condensation on the eggshell.

Once on a colder surface, like the eggshell, the steam condenses back into a liquid, and transfers the excess energy - the latent heat. It is that complex process which involves two phase transformations - vaporization and condensation - which allows us to cook eggs in the egg boiler within a reasonable preparation time. Note, the lid and the other internal parts of the device are all heated in the same way. Note also that the same heat transfer mechanism is at work when steaming any other food in the kitchen, e.g. vegetables.

**What happens inside the egg during heating?**

At its core, the process of boiling an egg, like many other methods of cooking food (for example, frying, baking, etc.), is based on the process of denaturation of proteins. At high temperatures, these complex organic molecules change their shape and spatial structure. Denaturation can be caused by many factors, including exposure to chemicals, enzymes, etc. [2,3,4]. The egg contains a mixture of about 100 proteins with different thermal properties. Unlike the physics of inorganic matter, where the exact value of the temperature of phase transitions (for example, melting or boiling) is known at a fixed pressure, for such complex organic compounds as proteins, the denaturation temperatures are different for each of the protein components in the yolk and eggwhite (they belong to the range *60-70 $^0C$*) .

The inhomogeneity of the temperature profile inside the egg during its heating is the main reason for the possibility to prepare a soft-boiled egg. The heat propagates from the eggshell inward gradually and heats the contents of the egg, denaturating the proteins on its way. The heat flow is proportional to the temperature difference between the shell and the centre of the egg. It was demonstrated that the temperature front propagates in the bulk of the egg following a square root law in time [1,2].

**The Paradox Solution**

In order to understand the origin of the "reciprocal law", we note that the water which re-condenses on the eggshells drips down and returns back to the heating plate. Then its evaporation-condensation cycle repeats over and over again. The more eggs, the larger the condensation surface, the larger the amount of water which is reused. Hence, cooking more eggs with the same initial amount of water will increase the rate of water recycling and therefore the total operation time. This will increase the exposure time of eggs to the steam. Therefore, they will be overcooked. Hence, in order to prepare a **larger number** of eggs cooked simultaneously to the correct point one has to use **less** water.

  As vapour condenses also at the internal surface of the lid, one recognizes that the law is not precisely reciprocal. Indeed, the marks on the cylindrical measuring cup are not equidistant. Some manufacturers produce the measuring cup in the form of a cone, making the marks approximately equidistant.

  So the main reason for the apparent paradox is that when we boil eggs to the desired degree of doneness, we must ensure the same cooking time for different numbers of eggs. It is this time that is the fundamental factor, and the initial amount of water is subordinate to it.

### Some estimates

Let us consider the case of steaming to the hard boiled condition[1] and assume that the egg has a mass of about *60 g*. The energy required to hard boil one egg is mainly determined by its heating from *5 $^0C$* (if initially kept in a fridge) to *100 $^0C$*. The average heat capacity of egg is about *3500 J/(kg K)*. The energy then amounts to *3500·95·0.06 J ≈ 20kJ*. This energy is obtained from the latent heat transfer in the process of steam condensation on the egg shell. The latent heat per 1g steam/water is *2264 J/g,* hence the mass of condensed water required is approximately *9 g*. This amount of water is returned back into the heating plate upon adding to boiler one more egg. If we want to keep the total steaming time constant, we must use *9 g* less water at the very beginning. This number agrees well enough with the measuring cup marks – the water amount reduction when adding one more egg for hard boiling is about *12g*.

### Measurements

We performed a number of measurements. When the boiler is switched on, a certain time $T_b$ is required to heat the initial amount of water to the boiling point. From then on, the actual steaming time $T_s$ is measured. The sum of the two is the processing time $T_p = T_b + T_s$. We started our experimentation with adding the amount of water prescribed for one hard-boiled egg, but running the device with no eggs, and without closing the lid. The evapourated water then leaves the device without condensing. The result amounts to $T_p$ = *973 s*, $T_b$ = *170 s*, $T_s$=*803 s* (Table 1). Next, we repeated the experiment after closing the lid, but still without eggs, noting that the inner surface of the lid acts as a condensation surface as well. We expect the processing and steaming times to increase. Indeed, we now measure $T_p$ = *1113 s*, an increase by 14% or *140 s* ($T_s$ = *943 s*). We then again repeated the experiment with both the lid and ONE egg, arriving at a processing time of nineteen minutes and forty eight seconds, $T_p$ = *1188 s*. The additional time increase by *75 s* is due to the additional condensation surface of one egg whose actual steaming time amounts to $T_s$ = *1018 s,* using the $T_b$ = *170 s* measurement obtained earlier .

| Lid | Eggs | Water for egg # | $T_p$ (s) | $T_b$ (s) | $T_s$ (s) |
| --- | --- | --- | --- | --- | --- |

---

[1] The cooking time $T_s$ of the egg prepared to a desired (say soft boiled) condition was derived by the British scientist Peter Barham and it is determined by the initial temperature $T_0$ and that one established at its surface $T_b$, egg's diameter *d*, intended final temperature of the yolk [5].

| | | | | | |
|---|---|---|---|---|---|
| off | 0 | 1 | 973 | 170 | 803 |
| on | 0 | 1 | 1113 | 170 | 943 |
| on | 1 | 1 | 1188 | 170 | 1018 |
| on | 3 | 3 | 1153 | 120 | 1033 |
| on | 7 | 7 | 1250 | 104 | 1146 |

Table 1. Times required to steam eggs to hard boiled condition

We kept increasing the number of eggs, the corresponding results are shown in Table 1. Overall, we observe that the steaming time for the hard-boiling condition does not substantially change. We note that $T_s$ is about 17 minutes and clearly larger than the time to boil an egg to the hard-boiled condition in boiling water (10-14 minutes). Yet the precise time to reach hard-boiled condition is somewhat blurred since there is a minimum time, but no sharp upper bound.

In order to assess the difference in steaming and boiling times, we repeat our egg steamer experiments for soft-boiled condition, which requires a much more precise and sharp preparation time. The results are shown in Table 2.

| Lid | Eggs | Water for egg # | $T_p$ / s | $T_b$ / s | $T_s$ / s |
|---|---|---|---|---|---|
| off | 0 | 1 | 227 | 45 | 182 |
| on | 0 | 1 | 350 | 45 | 305 |
| on | 1 | 1 | 470 | 45 | 425 |
| on | 3 | 3 | 480 | 40 | 440 |
| on | 7 | 7 | 600 | 35 | 565 |

Table 2. Times required to steaming eggs to soft boiled condition data.

The steaming time for one egg to soft-boiled condition amounts to $T_s = 425$ s. This is an increase of 18% as compared to the six minutes for boiling an egg to soft-boiled condition. The temperature difference between the egg centre (about 5 $^0C$) and the egg surface in boiling water (100 $^0C$) is 95 $^0C$. To transfer the same amount of heat (and therefore to reach the same preparation condition) during a longer steaming time implies that the egg surface during steaming is colder.

If the egg surface temperature changes (from boiling to steaming) then the temperature difference between surface and inner part changes. Using the Fourier law (heat flux is proportional to the temperature gradient, which can be approximated as the temperature difference between the outer and inner parts of the egg), the heat flux will decrease by the ratio of the two temperature differences. The steaming to boiling time change amounts to a factor of about

1.18. Therefore, the temperature difference while steaming will be *95 $^0C$ /1.18 = 80.5 $^0C$*. Adding *5.7 $^0C$* of the originally measured egg temperature from the fridge, we arrive at a surface temperature of the egg during steaming of *86.2 $^0C$*.

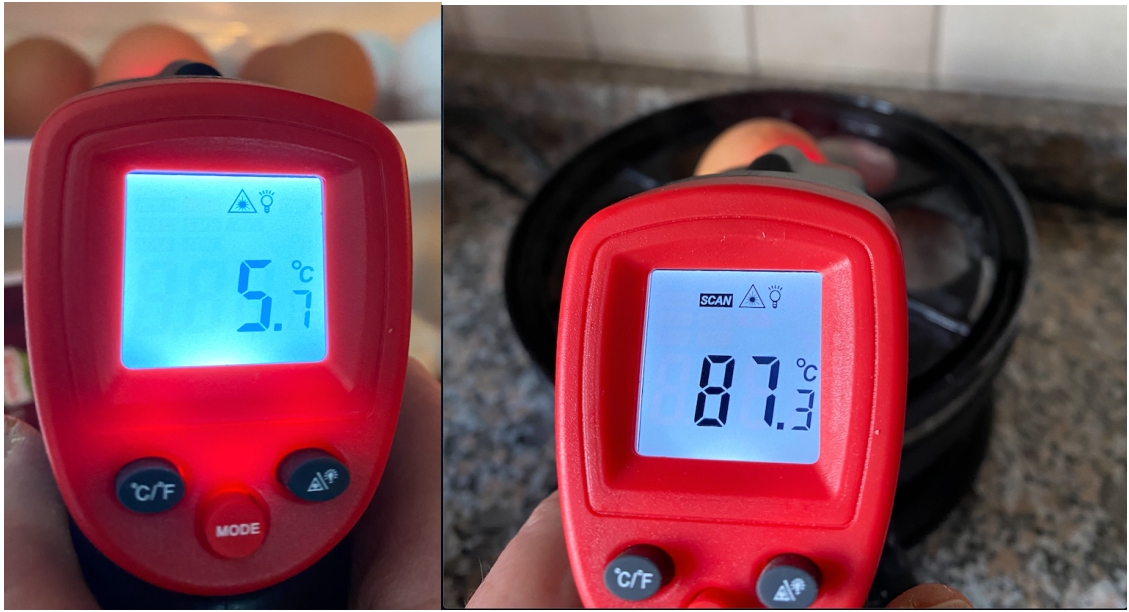

*Fig.2 The surface temperature of an egg before and after the steaming process, measured with an infrared thermometer.*

We measured the eggshell temperature during steaming using an infrared thermometer. We waited until the steaming process sets in for another two minutes. Then we removed the lid quickly and measured the eggshell temperature. The entire measurement takes less than a second. The result shown in Fig. 2 is *87.3 $^0C$*, which agrees reasonably well with our estimate of *86.2 $^0C$*, and our entire theory of the physics of egg steaming[2].

**Egg steamer efficiency**

Finally, we discuss the efficiency of the energy consumption during steaming. It takes about 20 minutes to boil hard-boiled eggs. The previously estimated amount of energy which is needed to hard-boil one egg amounted to 20kJ, which is equivalent to 1 minute of operation of a *350 W* heater. Steaming 6 eggs to the hard-boiled condition takes about 20 minutes. Thus, the efficiency of the device is about 25%.

**Conclusions**

---

[2] Let us note, that besides the accuracy of the thermometer itself (±1%), a considerable uncertainty is introduced by the measurement environment: namely, the fridge itself. We noticed that, depending on the degree of filling of the fridge by products, the measured temperature of the egg placed at the same place can change from *5.5 $^0C$* up to *10 $^0C$*.

We have discussed the physics of steaming eggs. We find that the main mechanism of heat transfer is the release of latent heat during condensation on the eggshell since the thermal conductivity by inelastic molecule collisions in the case under consideration is much less efficient. Increasing the number of eggs increases the condensation surface, and hence the volume of the condensed water returned to the heating plate, and consequently the processing time. To keep the latter constant, one needs to reduce the initial amount of water. We have calculated the required reduction of water volume when adding one egg (about 10g) which agrees well with the steamer device data. Based on processing time measurements, we have also estimated the egg surface temperature established during the steaming process. Our quantitative estimate was confirmed by direct infrared thermometer measurements. To repeat the experiments at home, one needs an egg steamer, a stopwatch, an infrared thermometer, and a box of eggs. The experiments can be easily performed at home without leaving the house, which is especially valuable during a pandemic.

The authors thank Dr. Charles Downing and Dr. Igor Smolyarenko for their carefully checking the manuscript and valuable comments.